\documentclass[11pt]{article}
\usepackage{authblk}
\usepackage{latexsym}
\usepackage{amssymb}
\usepackage{amsmath}
\usepackage{amsthm}
\usepackage{graphicx}
\usepackage{fullpage}
\usepackage{authblk}
\usepackage[hidelinks]{hyperref}
\usepackage{enumitem}
\usepackage[longend,noresetcount,boxed]{algorithm2e}
\usepackage{tikz}
\usetikzlibrary{positioning}
\usepackage{xcolor}
\usepackage{colortbl}
\usepackage{ifthen}
\usepackage{soul}
\usepackage{color}

\usepackage{anysize}
\usepackage{microtype}
\marginsize{1in}{1in}{1in}{1in}
\begin{document}
\DontPrintSemicolon
\LinesNotNumbered
\SetAlCapSkip{8pt}
\def\qed{\hfill$\Box$}
\def\Nset{\mathbb{N}}
\def\Ascr{\mathcal{A}}
\def\Bscr{\mathcal{B}}
\def\Cscr{\mathcal{C}}
\def\Dscr{\mathcal{D}}
\def\Escr{\mathcal{E}}
\def\Fscr{\mathcal{F}}
\def\Hscr{\mathcal{H}}
\def\Iscr{\mathcal{I}}
\def\Jscr{\mathcal{J}}
\def\Mscr{\mathcal{M}}
\def\Nscr{\mathcal{N}}
\def\Pscr{\mathcal{P}}
\def\Qscr{\mathcal{Q}}
\def\Rscr{\mathcal{R}}
\def\Sscr{\mathcal{S}}
\def\Uscr{\mathcal{U}}
\def\Wscr{\mathcal{W}}
\def\Xscr{\mathcal{X}}
\def\cupp{\stackrel{.}{\cup}}
\def\bold{\bf\boldmath}

\newcommand{\NP}{\mbox{\slshape NP}}
\newcommand{\opt}{\mbox{\scriptsize\rm OPT}}
\newcommand{\lin}{\mbox{\scriptsize\rm $OPT_{LP}$}}
\def\Sp{\mathop{\rm Sp}}
\def\cSp{\mathop{\rm cone.Sp}}
\newtheorem{theorem}{Theorem}
\newtheorem{lemma}[theorem]{Lemma}
\newtheorem{corollary}[theorem]{Corollary}
\newtheorem{proposition}[theorem]{Proposition}
\newtheorem{definition}[theorem]{Definition}
\newtheorem{observation}[theorem]{Observation}
\newtheorem{claim}{Claim}
\newcommand{\R}{\ensuremath{\mathbb{R}}}
\newcommand{\N}{\ensuremath{\mathbb{N}}}

\def\prove{\par \noindent \hbox{{\bf Proof}:}\quad}
\def\endproof{\eol \rightline{$\Box$} \par}
\renewcommand{\endproof}{\hspace*{\fill} {\boldmath $\Box$} \par \vskip0.5em}
\newcommand{\mathendproof}{\vskip-1.8em\hspace*{\fill} {\boldmath $\Box$} \par \vskip1.8em}
\newcommand{\E}[1]{\mathbb{E}\left[#1\right]}
\newenvironment{proof_claim}{\noindent {\it Proof of Claim: }}{\hspace*{\fill} $\diamond$\smallskip}

\title {Layers and Matroids for the Traveling Salesman's Paths \footnote{This research was partially supported by LabEx PERSYVAL-Lab (ANR 11-LABX-0025), by a grant from the Simons Foundation (\#359525, Anke Van Zuylen) and NSF grants CCF-1526067 and CCF-1522054.}}
\author{Frans Schalekamp}
\affil{\small School of Operations Research and Information Engineering\\
Department of Computer Science\\
Cornell~University\\
Ithaca NY, USA\\
fms9@cornell.edu
}
\author{Andr\'as Seb\H{o}}
\affil{\small Optimisation Combinatoire (G-SCOP)\\
CNRS, Univ. Grenoble Alpes\\
Grenoble, France\\
andras.sebo@grenoble-inp.fr
}
\author{Vera Traub}
\affil{\small Research Institute for Discrete Mathematics\\
University of Bonn\\
Bonn, Germany\\
traub@or.uni-bonn.de}
\author{Anke van Zuylen}
\affil{\small Department of Mathematics\\
College of William \& Mary\\
Williamsburg VA, USA\\
anke@wm.edu}

\maketitle

\begin{abstract}
Gottschalk and Vygen proved that every solution of the subtour 
elimination linear program for traveling salesman paths is a convex 
combination of more and more restrictive ``generalized Gao-trees''. We 
give a short proof of this fact, as a {\em layered} convex combination of 
bases of a sequence of increasingly restrictive matroids. A strongly 
polynomial, combinatorial algorithm follows for finding this convex 
combination, which is a new tool offering polyhedral insight, already 
instrumental in recent results for the $s-t$ path TSP.

\smallskip
\noindent{\footnotesize {\bf keywords}: path traveling salesman problem (TSP), matroid partition,
    approximation algorithm, spanning tree, Christofides' heuristic, 
    polyhedra}
\end{abstract}

\section{Introduction}
Gottschalk and Vygen~\cite{JC} proved that every solution
of the well-known  subtour elimination linear program for traveling salesman paths is a convex combination of a set of more and more restrictive ``generalized Gao  trees''  of the underlying graph, where a Gao-tree 
is a spanning tree that meets certain cuts in exactly one edge.

In this paper we provide  {\em layered} convex combinations of bases of a sequence of more and more restrictive matroids for a larger set of points, which we call chain-points, generalizing  the subtour elimination feasible solutions. This leads to a new connection of the TSP to matroids (observed in~\cite{AA}),  offering also a polyhedral insight, with specific algorithmic consequences, such as a strongly-polynomial combinatorial algorithm for finding this convex combination via the matroid partition theorem.  In this paper we show how the technical difficulties for proving the existence of such a particular convex combination and of turning it into an algorithm can be avoided, and a simple  proof follows. 

The existence of such a convex combination has been used by Seb\H{o} and Van Zuylen~\cite{AA} to prove the so far best upper bound on the integrality gap for the $s-t$ path traveling salesman problem. (We note that a recent result by Traub and Vygen~\cite{TV} gives an approximation ratio of $3/2 +\varepsilon$ for any constant $\varepsilon>0$, but this result does not imply an improved bound on the integrality gap.) This convex combination and the method we exhibit now to prove this result may possibly be adapted for proving further results on versions of the traveling salesman problem. The statement itself and its connection to matroids is one more link of the TSP to matroid partition or intersection, a connection interesting for its own sake.

In Section~\ref{sec:layers} we introduce chain-points, a key notion of our proof, and layered convex combinations, the related matroids, some preliminaries about these and some simple assertions that enable us to ``peel off'' layers one by one. In Section~\ref{sec:peel} we execute the induction step of peeling off a layer. 
We finish this introduction by introducing some notation and terminology.

We  assume throughout  that we are given a graph $G=(V,E)$, and $s,t\in V$.
For $S_1, S_2 \subseteq V$, we denote by $\delta(S_1,S_2) := \big\{ \{i,j\} \in E : i \in S_1, j \in S_2$\big\}. If $S_1 = S_2$ we use $E(S_1) = \delta( S_1, S_1 )$; if $S_1 = V \setminus S_2$ we  use $\delta( S_1 ) = \delta( S_2 ) = \delta( S_1, V \setminus S_1 )$. For $v \in V$, we let $\delta( v ) = \delta( \{ v \} )$. We denote by $G(S)=(S, E(S))$ the subgraph induced by $S\subseteq V$. For $F\subseteq E$ and $x\in \R^E$, we denote $x(F) := \sum_{e\in F} x_e$. With an abuse of notation and terminology, sets and their incidence vectors will not be distinguished. We denote by $\Sp(G)=\{x\in\mathbb{R}^E:x\ge 0, x(E(U))\le |U|-1\hbox{ for all } U\subseteq V,\, U\ne\emptyset,\hbox{ and } x(E)=|V|-1\}$ the spanning tree polytope (convex hull of spanning trees) of $G$, and by $\cSp(G)$ their cone (vectors that are their non-negative combinations);
$\Sp(G)\subseteq \cSp(G)$.

We say that $x \in \R^E$ {\em is a solution to the ($s-t$-path TSP) subtour elimination LP} if $x\in \Sp(G)$,  and $x$ satisfies the {\em degree constraints} $x( \delta(v) ) = 2$ for all $v \in V\setminus\{s,t\}$ and $x( \delta(s) ) = x( \delta(t) ) = 1$.
This is equivalent to the ``standard definition'' of the subtour elimination linear program: 
for $\emptyset \subset U\subseteq V\setminus\{s,t\}$, the constraints $x(\delta(U))\ge 2$ and  $x(E(U))\le |U|-1$  are equivalent if $x$ satisfies the degree constraints, and  for $\{s\} \subseteq U\subseteq V\setminus\{t\}$ similarly,  $x(\delta(U))\ge 1$ and $x(E(U))\le |U|-1$  are equivalent if $x$ satisfies the degree constraints.

\section{Chain-points, Layers and Matroids}\label{sec:layers}

Let $\emptyset \neq V_0 \subsetneq  V_1\subsetneq \ldots \subsetneq  V_k\subsetneq  V$, and $\Cscr:=\{\delta(V_0),\delta(V_1),\ldots,\delta(V_k)\}$. 
We call spanning trees of $G$ that meet each $C\in\Cscr$ in exactly one edge {\em Gao-trees} for the chain  
$V_0 \subsetneq  V_1\subsetneq \ldots \subsetneq  V_k$.
The {\em Gao-edges} of a cut $Q\in\Cscr$ are the edges $e\in Q$ for which $e\notin C$ for all $C\in \Cscr\setminus\{Q\}$.

We say   that $x\in\R^E$ is a {\em chain-point} for the chain of sets  
$V_0 \subsetneq  V_1\subsetneq \ldots \subsetneq  V_k$,
   if
   \begin{itemize}
      \item [(i)] $x\in\Sp(G)$,
      \item [(ii)] $x(\delta(V_0))=x(\delta(V_{k}))=1$,  $x(\delta(V_i))<2$ $(i=1,\ldots k-1)$,
      \item [(iii)] 
      $\sum_{v\in L_i}x(\delta(v))=2|L_i|$ $(i=1,\ldots k)$,  
   \end{itemize}
where $L_i:=V_i\setminus V_{i-1}\ne \emptyset$ $(i=0,\ldots, k+1)$ are the  {\em level-sets} (where $V_{-1}:=\emptyset$ and $V_{k+1}:=V$).  Note that levels $L_0$ and $L_{k+1}$ are not present in (iii).

Gao-edges can then be equivalently defined as  those joining two consecutive level-sets. 

When $V_0 \subsetneq V_1\subsetneq \ldots \subsetneq V_k$ are clear from the context, we use the terms chain-point and Gao-tree without mention of  the chain. Gao~\cite{Gao13} proved that there exists a Gao-tree in the support of a subtour elimination LP solution $x$. It is not hard to adapt Gao's proof to chain-points; it also follows from Lemma~\ref{lem:Gao} below (with $U=V$ and $E$ the support of $x$). 
Observe that Gao-trees are the bases of the matroid that is formed as the direct sum 
of the cycle matroids for the level sets $L_0, \dots, L_k$, whose bases are the spanning trees for the levels sets,
and the uniform matroids on the Gao-edges for each of the cuts $Q\in \Cscr$, whose bases are the one element subsets. 

As An, Kleinberg and Shmoys~\cite{AnKS12bis} noted, if $x$ is a solution to the $s-t$ path TSP subtour elimination LP, then {\em $\{V_0,V_1,\ldots,V_k\} := \{ A \subseteq V\setminus \{t\} : x(\delta(A)) <2 \text{ and } s\in A\}$ form a chain}, i.e., $V_0 \subsetneq V_1\subsetneq \ldots \subsetneq V_k\subsetneq  V$, where $V_0=\{s\}$ and $V_k = V\setminus \{t\}$. 
To check this for the sake of self-containedness,
let  $A,B\in \{V_0,V_1,\ldots,V_k\}$, and suppose for a contradiction that $A\setminus B\neq\emptyset$ and $B\setminus A \neq \emptyset$.
It will be convenient to consider $A$ and $C:= V\setminus B$. Observe that $x(\delta(A))+x(\delta(C)) = x(\delta(A)) + x(\delta(B)) < 2 + 2=4$, so we have \[4>x(\delta (A)) + x(\delta (C))= x(\delta (A\cap C)) +
x(\delta (A\cup C))+ 2 x(\delta(A\setminus C, C\setminus A))  \leqno{(\rm SUBMOD)}\]
by a well-known identity. Our assumption that $A\setminus B\neq\emptyset$ and $B\setminus A \neq \emptyset$ is equivalent to $A\cap C\neq \emptyset$ and $A\cup C\neq V$, where $A \cap C$ contains neither $s$ nor $t$ and $A\cup C$ contains both. Using  that $x$ is a solution of the $s-t$-path subtour elimination LP, 
$x(\delta(A\cap C))+ x(\delta(A\cup C))\ge 2+2 = 4$, a contradiction.

\medskip
{\em A solution $x$ to the $s-t$ path TSP subtour elimination LP is a chain-point for the chain  
$\{ A \subseteq V\setminus \{t\} : x(\delta(A)) <2 \text{ and } s\in A\}$,} 
since in this case $V_0 =\{s\}, V_k =V\setminus \{t\}$ 
and thus the conditions (ii) and (iii) are implied by the degree constraints.

Gottschalk and Vygen~\cite{JC} showed that it is possible to write a solution $x$ to the $s-t$ path TSP subtour elimination LP as 
a convex combination of spanning trees such that for every $\lambda \in [0,1]$ the coefficients of  Gao-trees for the chain $\{ A \subseteq V\setminus \{t\}  : x(\delta(A)) \leq 2-\lambda \text{ and } s\in A\}$ sum to at least $\lambda$. We call such a convex combination {\em layered} (see below for details).

The advantage of the notion of chain-points rather than subtour elimination LP solutions is that chain-points are closed under ``subtracting Gao-trees'' (while the set of feasible solutions to the subtour elimination linear program does not have this property due to the degree constraints).  This allows ``peeling off layers'' one by one and   proving the existence of a layered convex combination by induction (Lemma~\ref{lem:peeling} and Theorem~\ref{thm:main}).

We now define layered convex combinations somewhat more generally for chain-points, and in terms of the introduced matroids, in more detail, and state our main result, the existence of such a convex combination.  Then we state the lemma that allows to peel off layers one by one, that is, to deal with only two layers at a time. 

Fix a chain $\emptyset\neq V_0 \subsetneq V_1\subsetneq \ldots \subsetneq V_k\subsetneq V$. 
For a chain-point $x$, we call the values of $x(\delta(V_i))$ for $i=0,1,\ldots, k$ the {\em narrow cut sizes}. Let the different values of the narrow cut sizes be  
$2-\lambda_1 > 2-\lambda_1-\lambda_2 > \ldots > 2-\lambda_1-\ldots-\lambda_\ell=1$, where $\ell$ is the number of different sizes. 

We consider $\ell$ different matroids $(E, \Bscr_j)$ whose set of bases $\Bscr_j$ are the Gao-trees for the chain of sets $\{V_i:x(\delta(V_i))\le 2-\lambda_1-\ldots-\lambda_j\}$, $(j=1,\ldots, \ell).$ 
We say that $\sum_{j=1}^{\ell}\lambda_jx_j$ is a {\em layered convex combination} for $x$  if $x_j$ is in the convex hull of  bases in $\Bscr_j$ for $j=1, \ldots, \ell$ and $x=\sum_{j=1}^{\ell}\lambda_jx_j$.  

\begin{theorem}\label{thm:layered} If $x$ is a chain-point, then there exists a layered convex combination for $x$. 
\end{theorem}

The following lemma enables us to ``peel off'' layers of chain-points one by one, and concentrate on the case when there are only two distinct narrow cut sizes.

Denote the convex hull of Gao-trees of $G$ for the chain $\Cscr$ by $\Sp_\Cscr (G)$.
\begin{lemma}\label{lem:peeling}
	Let $x$ be a chain-point,  and  
	$x = \varepsilon y + (1-\varepsilon) x'$ with $y\in\Sp_\Cscr(G)$,  $x'\in \Sp(G)$, $\varepsilon \in [0,1)$. 
	Then $x'$ is a chain-point for the chain $(V_i: x(\delta(V_i)) < 2 - \varepsilon, i\in\{0,1, \dots, k \})$. 
\end{lemma}

\prove Since $y\in\Sp_\Cscr(G)$, for all $i=1,\ldots,k$ we have 
$y(\delta(L_i)) = 2$ and $y(E(L_i))= |L_i|-1$;
hence, $\sum_{v\in L_i} y(\delta(v)) = 2(|L_i|-1) + 2= 2|L_i|$.
 Since $x$ is a chain-point,  $\sum_{v\in L_i} x(\delta(v)) = 2|L_i|$ by (iii). 
 Now (i) holds for $x'$ by assumption; to check (ii) note $x'(\delta(V_i)) = \frac{x(\delta(V_i))-\varepsilon}{1-\varepsilon}$, whence $x'(\delta(V_0))=x'(\delta(V_k))=1$ and $x'(\delta(V_i)) < 2$ if $x(\delta(V_i))<2-\varepsilon$ ($i=0,\ldots, k$). 
  Finally, $\sum_{v\in L_i}x'(\delta(v))=\frac{2|L_i|-\varepsilon 2|L_i|}{1-\varepsilon}= 2|L_i|$ ($i=1,\dots, k$), 
  so (iii) also holds for $x'$.
\endproof 

\begin{theorem}~\label{thm:main}
	Let $x$ be a chain-point, and let $\lambda=2-\max_{i=0,\ldots, k} x(\delta(V_i))$. 
	Then there exist $y\in \Sp_\Cscr(G), x'\in \Sp(G)$ such that
	\[x=\lambda y + (1-\lambda) x'.\]
\end{theorem}

By Lemma~\ref{lem:peeling} applied to $\varepsilon:=\lambda$, the point $x'$ provided by this theorem is a chain-point for $(V_i: x(\delta(V_i)) <2-\lambda, i\in\{0,\dots,k\} )$, so by repeatedly applying Theorem~\ref{thm:main} we get Theorem~\ref{thm:layered}:

We prove this by induction on $\ell$, the number of different narrow cut sizes. 
Let the different values of the narrow cut sizes be $2-\sum_{j=1}^h\lambda_j$ for $h=1,\ldots, \ell$.
If $\ell=1$, then $\lambda_1=1$ by condition (ii), and thus $x\in \Sp_\Cscr(G)$, so $x$ is itself a layered convex combination for $x$.
If $\ell >1$, by Theorem~\ref{thm:main} there exists $y\in \Sp_\Cscr(G), x'\in \Sp(G)$ such that $x=\lambda_1 y + (1-\lambda_1) x'$.
Observe that if $x(\delta(V_i)) = 2-\sum_{j=1}^h \lambda_j$ for some $h\in \{1,\ldots, \ell\}$, then $x'(\delta(V_i)) = (2-\sum_{j=1}^h \lambda_j - \lambda_1)/(1-\lambda_1) = (2 - \sum_{j=2}^h \lambda_j)/(1-\lambda_1)$ ($i=0,\ldots, k$). By Lemma~\ref{lem:peeling}, $x'$ is a chain-point for the chain $(V_i: x(\delta(V_i)) <  2-\lambda_1, i\in \{0,1,\ldots, l\})$, which thus has narrow cut sizes $(2-\sum_{j=2}^h \lambda_j)/(1-\lambda_1)$ for $h=2,\ldots, \ell$.

Let $\lambda'_j = \lambda_{j+1}/(1-\lambda_1)$ for $j=1,\ldots, \ell-1$, then 
by the inductive hypothesis, there exist a layered convex combination for $x'$, i.e., $x'=\sum_{j=1}^{\ell-1} \lambda'_j x'_j$, where $x'_j$ is in the convex hull of bases in $\Bscr_{j+1}$. We thus have $x=\lambda_1 y + (1-\lambda_1) \sum_{j=1}^{\ell-1} \lambda'_j x'_j = \lambda_1 y + \sum_{j=2}^\ell \lambda_j x'_{j-1}$ where $y$ is in the convex hull of bases in $\Bscr_1$ and $x'_1,\ldots, x'_{\ell-1}$ are in the convex hull of bases in $\Bscr_{2},\ldots, \Bscr_\ell$ respectively. Hence, there exists a layered convex combination for $x$.

We prove Theorem~\ref{thm:main} using the following fractional version of Edmonds' matroid partition theorem~\cite{Edm68}, which can be easily  stated and proved for rational input from the well-known integer version  by multiplying with the denominators of the occurring numbers. We include here a reduction to an explicitly stated version in  the literature, a theorem on fractional polymatroids,  pointed out to us by Andr\'as Frank.

\newcommand{\Q}{\ensuremath{\mathbb{Q}}}

\begin{lemma}\label{lem:matroidunion}
Let $M_1$ and $M_2$ be matroids on the same element set $E$, 
and denote by $r_i$ the rank function of the matroid $M_i$ ($i=1,2$).
Let  $w \in \R^E_{\geq 0}$, $\lambda_1, \lambda_2 \in\R_{\geq 0}$
and let $P_i$ be the convex hull of the independent sets of $M_i$ ($i=1,2$). 
There exist $x_1 \in P_1$, $x_2 \in P_2$ such that $\lambda_1 x_1 + \lambda_2 x_2 = w$ 
if and only if 
$\lambda_1 r_1( X ) + \lambda_2 r_2( X ) \geq w( X )$ for all $X\subseteq E$.
\end{lemma}	
\prove
If $x_i\in P_i$ then $x_i(X)\le r_i(X)$, hence $w=\lambda_1 x_1 + \lambda_2 x_2$ with $x_1\in P_1, x_2\in P_2$ implies $w(X)= \lambda_1 x_1( X ) + \lambda_2 x_2( X )\le \lambda_1 r_1( X ) + \lambda_2 r_2( X )$.

To prove the reverse direction, apply \cite[Theorem~44.6]{Schrijver-book} to the submodular set-functions on $E$, $f_i:=\lambda_i r_i$   for which the conditions are satisfied, and let $w_i\in P_{f_i}=\lambda_i P_i$ $(i=1,2)$, where $P_f$ is the polymatroid associated with $f$.   

Our conditions $\lambda_1 r_1( X ) + \lambda_2 r_2( X ) \geq w( X )$ for all $X\subseteq E$ and $w\ge 0$   
express exactly that $w\in P_{f_1+f_2}$. So, by \cite[Theorem~44.6]{Schrijver-book} $w=w_1+w_2$, where $w_i\in P_{f_i}=\lambda_i P_i$ $(i=1,2)$. 
\endproof

\section{Peeling off a Layer}\label{sec:peel}

Our only debt now is to show that one layer can be peeled off, that is, to prove Theorem~\ref{thm:main}. 

We use Lemma~\ref{lem:matroidunion} with $w:= x$, $M_1$ the matroid whose bases are the Gao-trees and $M_2$ the cycle matroid of $G$, i.e., the matroid on $E$ whose independent sets are forests in $G$. We denote by  $p$ the rank function of $M_1$ and by $r$ the rank function of $M_2$. Clearly, $p\le r$. 
In the remainder of this section, we show that the condition of Lemma~\ref{lem:matroidunion} is satisfied for $\lambda_1=\lambda, \lambda_2=1-\lambda$, i.e. that 
$ \lambda p(X) + (1-\lambda) r(X) \ge x(X)$ for all $X\subseteq E$. Lemma~\ref{lem:matroidunion} then implies the existence of $y$ and $x'$ in the convex hull of independent sets of $M_1$ and $M_2$ respectively, and $x(E)= |V|-1$ implies that they are in fact in the convex hull of {\em bases} of $M_1$ and $M_2$, i.e., in $\Sp_\Cscr(G)$ and $\Sp(G)$ respectively, and Theorem~\ref{thm:main} follows.

To prove that the condition of Lemma~\ref{lem:matroidunion} is satisfied we need the following lemma.

\begin{lemma}\label{lem:Gao}
	Let $x$ be a chain-point and $U\subseteq V$ such that $x(E(U))=|U|-1$.  Then \[p(E(U))=r(E(U))=|U|-1.\]
\end{lemma}
\prove
We need to show that there exists a spanning tree $F$ in $G(U)$ that is an independent set of $M_1$, i.e., there exists a Gao-tree containing $F$. We begin by proving two claims.

\begin{claim}\label{prop1} The set $I(U):=\{i\in[0,k+1]: L_i\cap U\ne\emptyset\}$ is the set of all integers of an interval.
\end{claim}

Indeed, suppose for a contradiction that there exists $j\in [0,k]$, $L_j\cap U=\emptyset$ such that $L_{j'}\cap U\ne\emptyset$ for some $j'<j$ and also for some $j'>j$; in other words, $U$ is partitioned by $U\cap V_{j-1}$ and $U\setminus V_j$.

By (ii), and applying (SUBMOD) for $A:=V_j,C:=V\setminus V_{j-1}$, we get that:
\[4 > x(\delta(V_{j} ))+ x(\delta(V\setminus V_{j-1} )) = x(\delta(L_{j} ))+ x(\delta(V)) + 2 x( \delta( V_{j-1},V\setminus V_{j} ) ).\]
To derive the desired contradiction, we will show that the right hand side is at least 4.
First, $x(\delta(V))=0$, and (iii) implies that $x(\delta(L_j)) = 2|L_j| - 2x(E(L_j))$ which is at least 2 since $x(E(L_j))\le |L_j|-1$ for $x\in \Sp(G)$. Further, observe that
\begin{align*}
x( \delta( V_{j-1},V\setminus V_{j} ) )& \ge x(E(U)) - x(E(U\cap V_{j-1})) -  x(E(U\cap (V\setminus V_{j}))) \\
&\ge |U|-1- (|U\cap V_{j-1}|-1) - (|U\setminus V_{j}|-1)= 1,
\end{align*}
where the second inequality uses the fact that $x(E(U))=|U|-1$ and $x(E(A))\le |A|-1$ for any $A$, and the equality uses the fact that $\{U\cap V_{j-1}, U\setminus V_{j}\}$ is a partition  of $U$. The claim is proved.

The key claim we need  now to prove the lemma is the following.
\begin{claim} \label{prop2}
	Let $a,b\in\N,$ $0\le a \le b \le k+1$, and $S:=\bigcup_{i=a}^bL_i$. Then $E(S \cap U)$ is connected.
\end{claim}
Note that this was shown by Gao~\cite{Gao13} for $U:=V$ and is the key for the existence of a Gao-tree.

To prove Claim~\ref{prop2}, we first show that
\[
	|S|-2 < x(E(S)) \quad(\le |S|-1).\leqno{(\rm STRICT)}
	\]
If $a=0$, $b=k+1$, then $S=V$ and by  (i), $x(E(V))=|V|-1>|V|-2$; if exactly one of  $a=0$ or $b=k+1$ holds, then, slightly more generally, subtracting from $x(E(V))=|V|-1$ the inequality $x(E(V\setminus S))\le |V\setminus S|-1$  we get $x(E(S))+ x(\delta(S))\ge |S|$ (again from (i)), and then by  (ii) $x(\delta(S))<2$, and we are done again.  
Finally, suppose $1\le a\le b\le k$, and add up   (iii) for  $i=a,a+1,\ldots,b$. We get $x(E(S))+ \frac12x(\delta(S))=|S|$,
 where $\frac12x(\delta(S))\le\frac12x(\delta(V_{a-1}))+ \frac12x(\delta(V_{b}))<2$ by (ii), finishing the proof of (STRICT).

Now, by (i), $x$ is a convex combination of spanning trees, and by (STRICT), at least one of these, denote it $F$, contains a spanning tree of $S$; $F$, as all the spanning trees in the convex combination, also contains a spanning tree of $U$, because of $x(E(U))=|U|-1$.  So $F$ contains a spanning tree of $S\cap U$,  and the claim is proved.

\smallskip
To finish the proof of the lemma, 
choose a spanning tree in $G(L_i\cap U)$ for each $i\in I(U)$, which is possible by 
Claim~\ref{prop2} applied to $S:=L_i$; then add a Gao-edge between $L_i\cap U$ and $L_{i+1}\cap U$ for each index $i$ such that  $i,i+1\in I(U)$, which is possible by applying Claim~\ref{prop2} to $S:=L_i\cup L_{i+1}$. In this way we get a spanning tree $F$ of $G(U)$, which is an independent set of $M_1$, since it can be completed to a Gao-tree of $G$ by applying Claim~2 to $U:=V$ and $S:=L_i$ for $i=0,\ldots, k+1$ and then for $S:=L_i\cup L_{i+1}$  for $i=0,\ldots, k$, if $\{i,i+1\}\setminus I(U)\neq \emptyset$.
\qed

\bigskip
\noindent{\bf Proof of Theorem ~\ref{thm:main}}: 
First, observe that it suffices to prove the following claim. 
\begin{claim}\label{claim:eps}
There exists $0<\varepsilon \le \lambda$ and $y\in \Sp_\Cscr(G)$, $x'\in\Sp(G)$, such that
$x=\varepsilon y + (1-\varepsilon) x'.$
\end{claim}
Indeed, if this is true, then $x-\varepsilon y \in \cSp(G)$, and let $\varepsilon_{\max}$ be the largest $\varepsilon\le \lambda$ such that there exists $y\in \Sp_\Cscr(G)$ such that $x-\varepsilon y \in \cSp(G)$. 
Note that the $\varepsilon$ is well-defined, because 
$\{ (\varepsilon , y) : y\in \Sp_\Cscr(G),x-\varepsilon y \in \cSp(G), 0\le \varepsilon\le \lambda\}$ is a polytope.

Defining $x':=\frac{x - \varepsilon_{\max} y}{1-\varepsilon_{\max}}$, we see that $x'\in \Sp(G)$. So to prove Theorem~\ref{thm:main} from Claim~\ref{claim:eps} we have to prove $\varepsilon_{\max} = \lambda$.

Suppose for a contradiction that $\varepsilon_{\max} < \lambda$, then by Lemma~\ref{lem:peeling} applied to $\varepsilon:=\varepsilon_{\max}$, $x'$ is also a chain-point (for the same chain of sets).
So then applying Claim~\ref{claim:eps} to $x'$, there exists $\varepsilon'>0$ and $y'\in \Sp_\Cscr(G)$ such that $x'-\varepsilon' y' \in \cSp(G)$.
However, then with $\varepsilon'':=(1-\varepsilon_{\max})\varepsilon'$ we have
\[\cSp(G)\ni (1-\varepsilon_{\max})(x'-\varepsilon'y')= x- \varepsilon_{\max}y - \varepsilon''y'=x - (\varepsilon_{\max} + \varepsilon'')z,\]
where $z:=\frac{\varepsilon_{\max}y + \varepsilon''y'}{\varepsilon_{\max} + \varepsilon''}\in \Sp_{\Cscr}(G)$.
This contradicts that $\varepsilon_{\max}$ is the largest $\varepsilon\le \lambda$ such that there exist $y\in \Sp_\Cscr(G)$ such that $x'=x-\varepsilon y \in \cSp(G)$, finishing the proof of the theorem, provided that Claim~\ref{claim:eps} is true.

We prove now Claim~\ref{claim:eps} by checking the condition of the matroid partition theorem in the form of Lemma~\ref{lem:matroidunion}. Let us say that
$\varepsilon\ge 0$ is {\em suitable} for $X\subseteq E$,  if $\varepsilon>0$ and
\[\varepsilon p(X) + (1-\varepsilon) r(X) \ge x(X).\]
By Lemma~\ref{lem:matroidunion}, Claim~\ref{claim:eps} is equivalent to proving that {\em there exists $\varepsilon > 0$  suitable {\em for all} $X\subseteq E$.}

If $p(X)=r(X)$ then by (i) any $\varepsilon\in [0,1]$
is suitable for $X$; so assume $p(X) < r(X)$.
If in addition $x(X)< r(X)$, then $\varepsilon$ is suitable if and only if $0<\varepsilon\le\varepsilon_X:=\frac{r(X) - x(X)}{r(X)-p(X)}\, (>0)$. Since the number of such sets is finite, $\varepsilon_{\max}:=\min \{\varepsilon_X : X\subseteq E, x(X)< r(X), p(X)< r(X) \} >0$ is suitable for all of them.

It remains to show that  {\em  there is no other case}, that is, $x(X)=r(X)$ implies $p(X)=r(X)$. 
Denote $V_X$ the set of vertices of (``covered by'') edges in $X$. We consider the components of $(V_X, X)$.
If $V_X$ is empty, $X=\emptyset$ and hence $r(X)= 0 = p(X)$.
If $(V_X, X)$ has only one component, then $r(X)=|V_X|-1$. 
Apply Lemma~\ref{lem:Gao} to the graph whose edge-set is the support of $x$, and $U:=V_X$; then $X=E(U)$ and by the lemma, $p(X)=r(X)$ follows.

If $(V_X, X)$ has multiple components, then $r(X)$ sums up over the components. The same holds for $p(X)$ as long as there are no two components that both contain edges in the same cut in ${\cal C}$. (This can be easily seen from the definition in Section~\ref{sec:layers} of the matroid in terms of a direct sum of cycle matroids on the level sets and uniform matroids on the Gao-edges of each cut.)
Hence, the proof of the claim is completed by showing that it is not possible for two components $A$ and $B$ of $(V_X,X)$ that both $E(A)\cap \delta(V_i)\ne\emptyset$ and $E(B)\cap \delta(V_i)\ne\emptyset$.

Indeed, if $E(A)\cap \delta(V_i)\neq \emptyset$, then $x(E(A)\cap \delta(V_i)) = x(E(A)) - x(E(A\cap V_i)) - x(E(A \setminus V_i))\ge 1$,
since  $x(E(A))=|A|-1$ because of $x(X)=r(X)$, and  $x(E(A'))\le |A'|-1$ for any 
$\emptyset \neq A'\subseteq A$, because $x\in \Sp(G)$.  
By applying the same argument to $B$, if both $E(A)\cap \delta(V_i)\neq\emptyset$ and $E(B)\cap \delta(V_i)\neq \emptyset$, then $x(\delta(V_i)) \ge 2$, contradicting (ii).
\qed

\smallskip
Our proof implies that a layered convex combination can be  found in strongly polynomial time with a combinatorial algorithm. 
One way of achieving this is through Edmonds' matroid partition algorithm. The best-known fractionally weighted implementation of this is by Cunningham \cite{Cu84}, which can be modified for different matroids. 
Note that for rational weights the multiplication with a common denominator allows a direct application of matroid union, that can be implemented in strongly polynomial time. The rank oracle of the occurring matroids is at hand from the above proofs.

Another way of achieving this is by using the algorithm for writing a point $x$ in $\Sp(G)$ as a convex combination of spanning trees in the proof of Theorem 51.5 of~\cite{Schrijver-book}. At every iteration of this algorithm, a spanning tree $T$ is chosen that has maximum weight for a certain weight function. The weight function is determined by the algorithm; it satisfies the property that a tree $T$ is a maximum weight tree if and only if it satisfies $T(U)=|U|-1$ for certain sets $U$ for which $x(E(U))=|U|-1$. Given $T$, a value $\lambda\ge 0$ is computed so that $x' = (x-\lambda T)/(1-\lambda) \in \Sp(G)$. The algorithm then recurses on $x'$. Now, instead of choosing $T$ to be an arbitrary maximum weight spanning tree, we choose a maximum weight spanning tree that is a Gao-tree for the chain $(V_i: x(\delta(V_i)) < 2, i\in \{1,\ldots, k\})$. It follows from the results above that a Gao-tree with this property indeed exists.

It then follows from~\cite{Schrijver-book} that there exists a layered convex combination for $x$ for which the number of spanning trees is at most linear in the size of the support of $x$, which is the same as for an arbitrary decomposition into spanning trees. If $x$ is an extreme point solution of the subtour elimination linear program the support of $x$ has size $O(|V|)$~(see~\cite{Goemans06}); so only $O(|V|)$ different trees are needed for a layered convex combination.

\subsubsection*{Acknowledgment} 
Many thanks are due to  Andr\'as Frank, Zoli Szigeti, David Williamson and an anonymous referee for careful reading and helpful suggestions improving the presentation.

\end{document}